\documentclass[useAMS,usenatbib,usegraphicx]{mn2e}

\newcommand{\he}[1] {He\,{\sc #1}}
\newcommand{\hel}[2] {He\,{\sc #1}~$\lambda$#2}

\newcommand{\Ha}{\mbox{${\mathrm H\alpha}$}}
\newcommand{\Hb}{\mbox{${\mathrm H\beta}$}}
\newcommand{\Hg}{\mbox{${\mathrm H\gamma}$}}

\title[Circular polarisation in RX\,J1643.7+3402]{The magnetic SW Sextantis star RX\,J1643.7+3402}
\author[P. Rodr\'\i guez-Gil et al.]{P. Rodr\'\i guez-Gil$^{1,2}$\thanks{E-mail:
prguez@ing.iac.es; igm@iac.es; jaime@astro.su.se}, I. G. Mart\'\i nez-Pais$^{2,3}$ and J. de la Cruz Rodr\'\i guez$^{2}$\thanks{Present address: Institute for Solar Physics, Royal Swedish Academy of Sciences, AlbaNova University Center, SE-106 91 Stockholm, Sweden}\\
$^{1}$Isaac Newton Group of Telescopes, Apdo. de Correos 321, E--38700, Santa Cruz de la Palma, Spain\\
$^{2}$Instituto de Astrof\'\i sica de Canarias, C/ Via L\'actea s/n, E--38200 La Laguna, Tenerife, Spain\\
$^{3}$Departamento de Astrof\'\i sica, Universidad de La Laguna, E--38206 La Laguna, Tenerife, Spain}
\begin{document}

\date{Accepted 2009. Received 2008; in original form 2008}

\pagerange{\pageref{firstpage}--\pageref{lastpage}} \pubyear{2009}

\maketitle

\label{firstpage}

\begin{abstract}
We present time-resolved spectroscopy and circular spectropolarimetry of the SW Sex star RX\,J1643.7+3402. We find significant polarisation levels exhibiting a variability at a period of $19.38 \pm 0.39$ min. In addition, emission-line flaring is found predominantly at twice the polarimetric period. These two findings are strong evidences in favour of the presence of a magnetic white dwarf in the system. We interpret the measured periodicities in the context of our magnetic accretion model for SW Sex stars. In contrast with LS Pegasi ~---the first SW Sex star discovered to have modulated circular polarisation---~ the polarisation in RX\,J1643.7+3402 is suggested to vary at $2(\omega-\Omega)$, while the emission lines flare at $(\omega-\Omega)$. However, a $2\omega$/$\omega$ interpretation cannot be ruled out. Together with LS Peg and V795 Her, RX\,J1643.7+3402 is the third SW Sex star known to exhibit modulated circular polarisation.
\end{abstract}


\section[]{Introduction}

The ROSAT X-ray source RX\,J1643.7+3402 (AAVSO 1640+34) was identified as a cataclysmic variable (CV) by 
\citet{micka}. Based on spectroscopic and photometric data they classified this object as a member of the SW\,Sex family. In spite of their effort to determine its orbital period, two possible candidates from the power spectra of the \Hb~radial velocity curves, namely 2.575\,h and 2.885\,h, were suggested. In addition, they found a photometric period of 2.595\,h which they identified 
as a superhump modulation, as well as rapid variations in the $V$-band light curve with a time scale of $\sim 15$\,min.

In the same year, \citet{patterson-etal2002} measured a reliable value for the orbital period, 2.89344(34)\,h, in agreement with one of the values suggested by \citet{micka}. This places the system within the period gap, a region in the period distribution of CVs (between approximately 2 and 3\,h) in which a dearth of systems is observed. \citeauthor{patterson-etal2002} also reported two additional periodicities found in the light curve of RX\,J1643.7+3402: 2.807\,h and 4.05\,d, which they interpret as a negative superhump and the wobble 
period of the accretion disc, respectively. Moreover, a quasi-periodic oscillation (QPO) with a period near 17\,min (very close to that previously reported by \citeauthor{micka}) was also observed. On the other hand, \citet{patterson-etal2002} confirmed the SW\,Sex membership of RX\,J1643.7+3402 on the basis of its spectroscopic behaviour and photometric variability.

The SW\,Sex stars form a group of CVs firstly defined by \citet{thorstensen}. Their original definition was later updated at the light of new observational and theoretical knowledge (see e.g. Mart\'\i nez-Pais, Rodr\'\i guez-Gil \& Casares 1999, and \citealt{prguez2007b} for a review). Briefly, (i) they show a strong absorption component in the Balmer and \he{i} lines usually detectable along a fraction of the orbit and reaching maximum depth around orbital phase $\sim 0.5$, when it moves from positive to negative velocities. The relative intensity of this component increases with the line excitation level, even dominating the line profile in \he{i}; (ii) they also show a high-velocity emission S-wave which is stronger in the blue wing of the Balmer lines 
and reaches its extreme negative velocity at orbital phase $\sim 0.4-0.5$. (iii) As a consequence of the complex structure of the emission lines, the Balmer radial velocity curves are delayed with respect to the white dwarf motion. (iv) Most of them \citep[68 percent, see][]{prguez2007b} cluster at the upper edge of the period gap with orbital periods in the range $\sim 3-4$\,h.

The magnetic origin of the SW\,Sex phenomenology has been claimed by several authors (\citealt{williams89, casares96, mpais99}). It was only after the discovery of variable circular polarisation in the non-eclipsing systems LS\,Peg \citep{prguez2001} and V795\,Her \citep{prguez2002b}, and the X-ray pulsation in LS\,Peg at exactly the polarimetric frequency (Baskill, Wheatley \& Osborne \citeyear{baskilletal}), that the presence of a magnetic white dwarf has become a primary ingredient in defining the nature of this class of CV.

RX\,J1643.7+3402 (hereafter J1643) is the brightest non-eclipsing SW\,Sex star known (and one of the brightest CVs at $V \simeq 12.6$). This, together with the restrictive signal requirements of time-resolved spectropolarimetry, make it the best candidate for probing the presence of a magnetic white dwarf by means of circular polarimetry measurements. Besides, the lack of eclipses ensures an inclination angle small enough to guarantee a good visibility of the white dwarf along the whole orbital cycle. For this reasons we decided to undertake a spectropolarimetric study on this system similar to our pioneering work on LS\,Peg \citep[][hereafter P01]{prguez2001}, which provided a positive result: the detection of variable circular polarisation at a period of 29.6 ($\pm 1.8$) min.


\section[]{Observations and data reduction}

Two data sets (spectroscopic and spectropolarimetric) are presented in this paper. Both were obtained using the blue arm of the Intermediate dispersion Spectrograph and Imaging System (ISIS) mounted on the 4.2\,m William Herschel Telescope at the Roque de los Muchachos Observatory on La Palma. Spectra of Cu-Ne-Ar comparison lamps were obtained every $\sim 30-40$\,min in order to secure an accurate wavelength calibration. Data reduction was performed using {\sc iraf}\footnote{{\sc iraf} is distributed by National Optical Astronomy Observatories}, while wavelength calibration and most of the subsequent analyses made use of Tom Marsh's {\sc molly}\footnote{{\sc molly} is available at Tom Marsh' web page: {\tt http://deneb.astro.warwick.ac.uk/phsaap/software/}} package.

Regarding the pure spectroscopic data, a total of 114 spectra were acquired on 2002 May 21 using the R600B grating and a 1-arcsec slit width. This gave a spectral resolution of 1.7 \AA~(full-width at half maximum, FWHM) in the $\lambda\lambda3600-5350$ spectral range. The data were reduced in a standard way, i.e. by applying bias and flat-field 
corrections, sky subtraction and the optimal extraction procedure described by \citet{horne86}. Once the spectra were wavelength calibrated, their wavelength scale were re-sampled into an uniform velocity scale and then normalised to their continua. 

The spectropolarimetric data (2003 July 1) were acquired using a $\lambda$/4 plate above the slit and a Savart plate below it. This configuration produces two different images on the detector, corresponding to the ordinary and the extraordinary ray, respectively. The exposures were performed alternating two positions of the $\lambda$/4 plate differing by 90\degr. A total of 44 images were acquired at each position angle of the quarterwave plate, each lasting between 120 and 150\,s. We used the R300B grating which covered the spectral range $\lambda\lambda3800-7430$ at a 3.4\,\AA~resolution). Reduction of the spectropolarimetric data was performed according 
to the procedure described in P01. The extraction of the ordinary and extraordinary spectra from the images taken at both positions of the $\lambda/4$ plate provided a total of $4 \times 44=176$ spectra, from which the spectral density of the $P_V=V/I$ Stokes parameter can be obtained following the procedure described in \citet{tin-rut}. 

In addition to the spectropolarimetric information contained in the 2003 data it is possible to recover the (non-calibrated) flux distribution by just co--adding the ordinary and extraordinary spectra on each frame. We therefore obtained a total of 88 spectra which were later re-binned and rectified in the same way as the 2002 data.


\section[]{Variable circular polarisation}

Once the $P_\mathrm{V}$--spectra were obtained (see Sect.~2) we measured the circular polarisation level on different continuum windows. $P_\mathrm{V}$--HJD curves were subsequently constructed for which Scargle periodograms were computed. The best results (i.e. the 
strongest peaks in the periodograms) were found for redder continuum regions, especially in the range $\lambda\lambda$5960--6850 after 
masking out the atmospheric and interstellar features. It must be noted, however, that no polarisation was detected in the emission lines. Actually, 
neither \Ha~nor \hel{i}{6678} (the lines included in the working range) are present in the $P_\mathrm{V}$ spectra. 

\begin{figure}
\includegraphics[width=85mm]{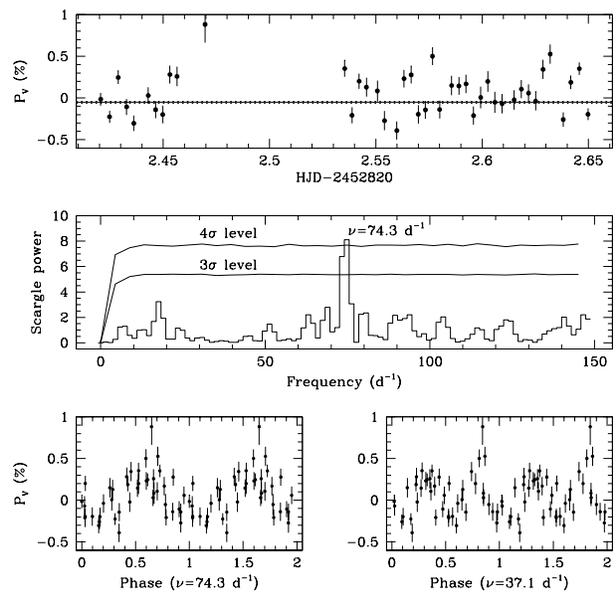}
\caption{{\em Upper panel}: continuum circular polarisation ($P_\mathrm{V}$) curve for the $\lambda \lambda$5960-6850 wavelength range. 
Also shown is the mean polarisation level obtained for the zero-polarisation star HD\,212311 \citep{turnshek}. The Scargle periodogram ({\em middle panel}) clearly 
shows a sharp peak centred at 74.3\,d$^{-1}$ exceeding the $4\sigma$ (99.99 per cent) significance level. The data folded on this frequency and 
half this frequency are also shown ({\em bottom panels}; note that a whole cycle has been repeated for clarity). The phase origin corresponds to 
the first data point.}
\label{fig:polarizacion}
\end{figure}

\begin{figure*}
\includegraphics[width=6.3cm,angle=-90]{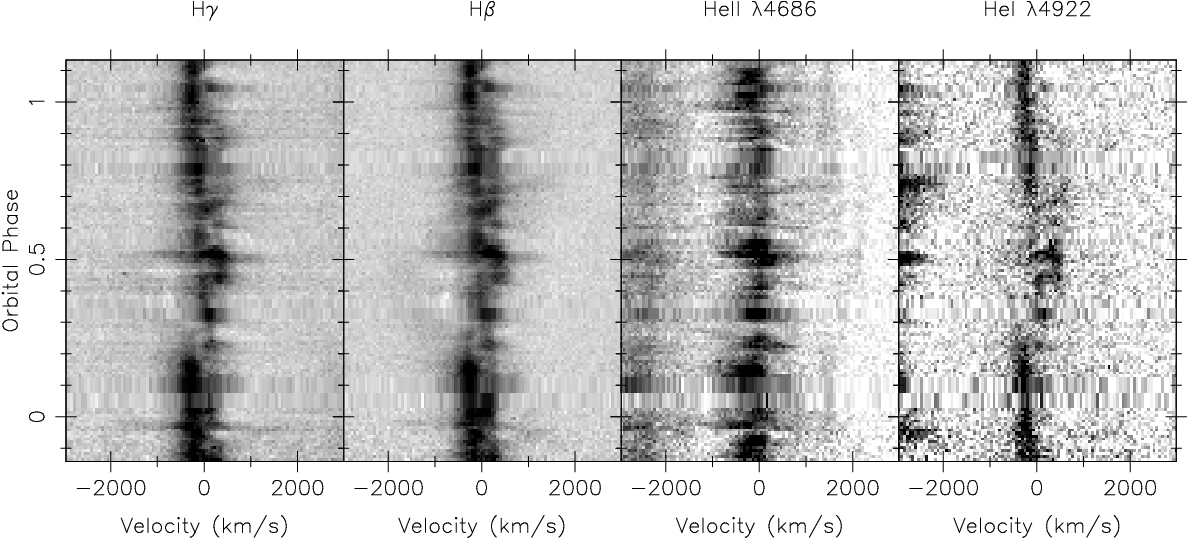}\\
~\\
\includegraphics[width=6.3cm,angle=-90]{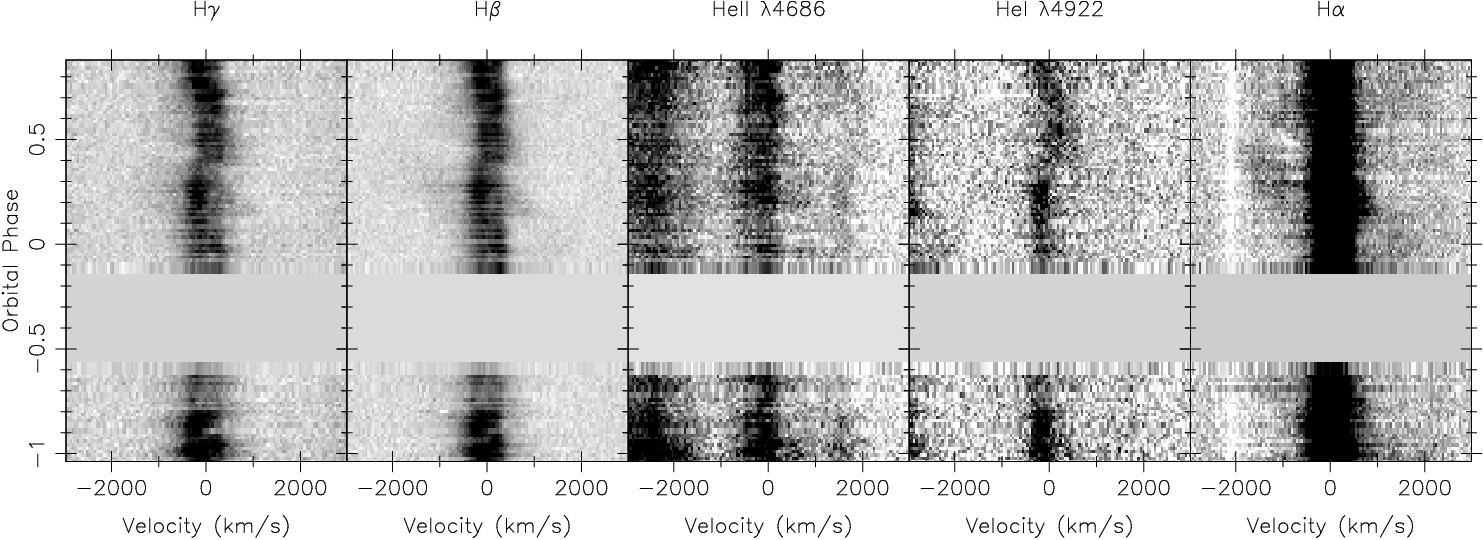}
\caption{{\sl Top}: Continuum-subtracted, trailed spectra diagrams in the vicinity of \Hg, \Hb, \hel{ii}{4686}, and \hel{i}{4922} from the May 2002 data. Emission-line flaring is apparent in all the lines, especially in \hel{ii}{4686}. Black represents emission. {\sl Bottom}: same for the July 2003 data but including \Ha. The gray levels of the \Ha~trailed spectra have been adjusted to enhance the high-velocity S-wave. The blank space corresponds to a bad weather interval. Orbital phases for both sets of data were calculated from the orbital period reported by \citet{patterson-etal2002} under the assumption of maximum blue excursion of the high-velocity S-wave occurring at $\varphi = 0.45$ \citep[see e.g.][]{prguez2007b}.}
\label{fig:trailed}
\end{figure*}

In Fig.~\ref{fig:polarizacion} we show the circular polarisation curve for the selected continuum region. The Scargle periodogram \citep{scargle} is clearly dominated by a sharp peak at $\sim 75$\,d$^{-1}$ for which a Gaussian fit 
yielded a frequency of $74.3 \pm 1.5$\,d$^{-1}$ (a period of $19.38 \pm 0.39$ min; the quoted uncertainty is half the FWHM of the fitting 
Gaussian). In order to check the significance of this periodicity we performed a Monte Carlo test by generating $10^6$ white-noise time series, i.e. random, zero-mean normal distributions of data with the same sampling as our $P_\mathrm{V}$ data. 
A Scargle power spectrum was subsequently computed for each white-noise time series and an statistic was performed among them. Thus, for 
example, the $4\sigma$ level indicated in Fig.~\ref{fig:polarizacion} is the power level for which 99.99 per cent of the computed 
power spectra have values below it. We can therefore say that the probability of a random time series having a peak in its periodogram 
exceeding this level is 0.01 per cent or, in other words, a peak having a power greater than this value has a probability of at least 
99.99 per cent of being true. This unambiguous detection of variable circular polarisation in J1643 is a clear sign of magnetic accretion in the system. Fig.~\ref{fig:polarizacion} shows the circular polarisation data folded on both the 19.38-min period and twice that value. The reason for this will be explained in what follows.


\begin{figure*}
\includegraphics[width=85mm]{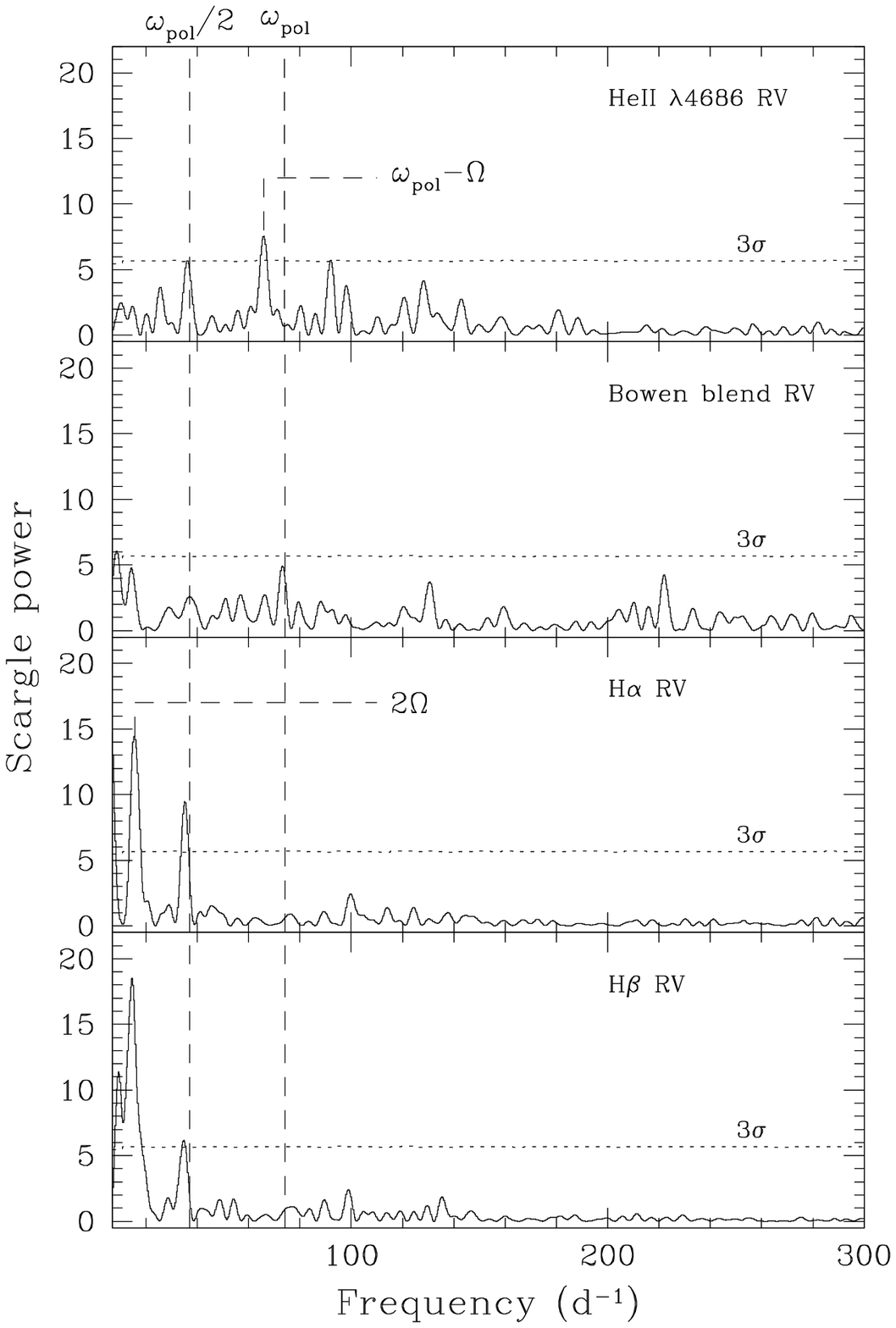} \includegraphics[width=85mm]{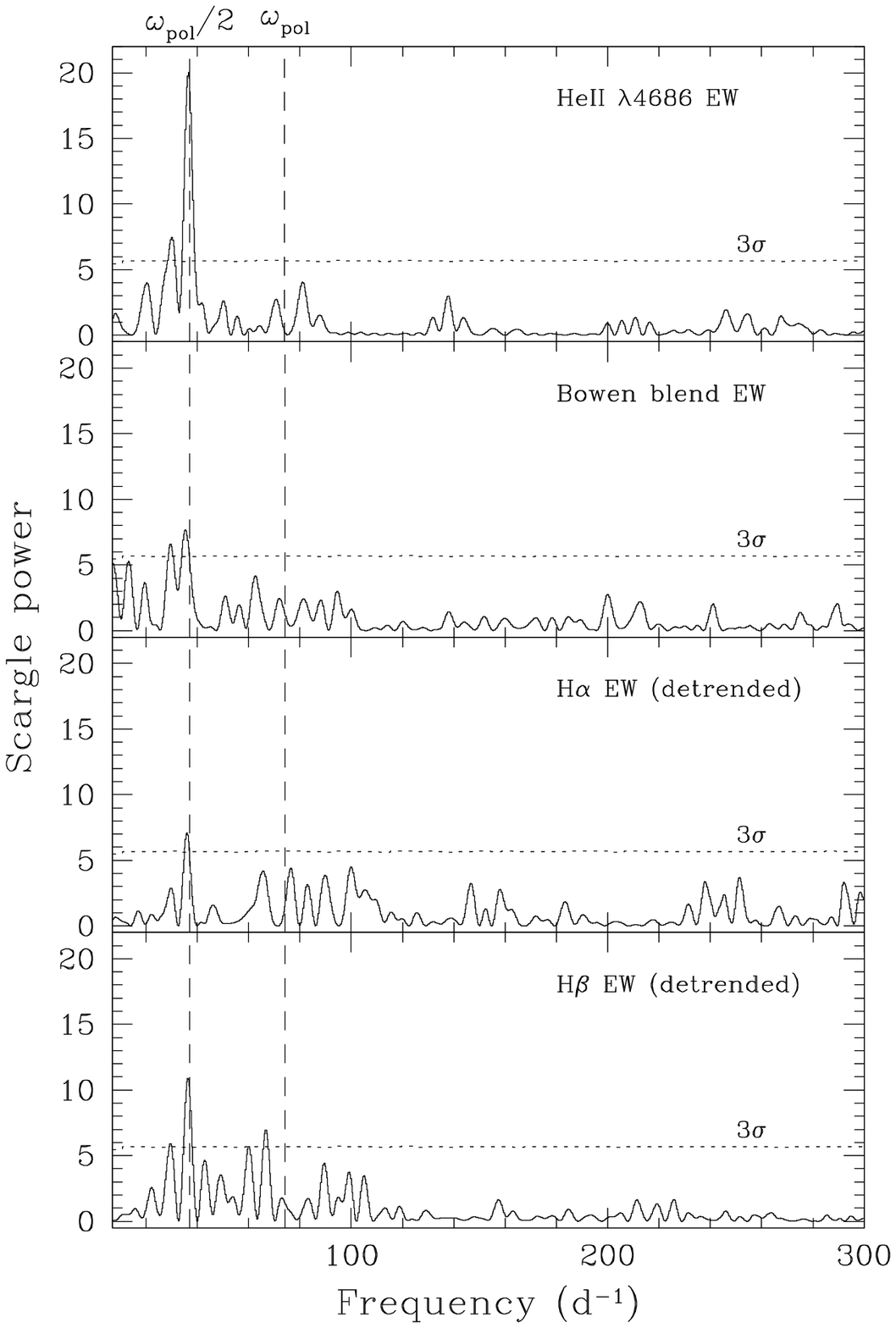}
\caption{{\sl Left panel}: Scargle periodograms of the 2003 \hel{ii}{4686}, Bowen blend, \Ha, and \Hb~radial velocity curves. {\sl Right panel}: same for the EW curves.}
\label{fig:rvel_ew_scargle_2003}
\end{figure*}

\section[]{Emission-line variability}

Magnetic accretion can give rise to variations of the emission lines at a period related with that of the circular polarisation variability. This is the case in e.g. the intermediate polar CV (IP) PQ Gem \citep{piirolaetal93-1, rosenetal93-1, potteretal97-1}, and the SW Sex star LS Peg (P01). In Fig.~\ref{fig:trailed} we present the \Hg, \Hb, \hel{ii}{4686}, \hel{i}{4922}, and \Ha~trailed spectra diagrams constructed from the pure spectroscopic data (May 2002) and the spectropolarimetric data (July 2003) after subtracting the continuum. Emission-line flaring at a short time scale ($\sim 10-20$\,min) is clearly visible, especially in the 2002 data. Apart from LS Peg, these variations have been observed in the emission lines of other four SW\,Sex stars \citep[see ][and references therein]{prguez2007b}, and are also seen in the optical spectra of other IPs, such as FO\,Aqr \citep{marsh+duck96}. By analogy to IPs, the emission-line flaring present in the SW\,Sex stars has been connected to the spin period of a magnetic white dwarf \citep[P01,][]{patterson-etal2002}.

In order to search for coherent signals in the lines of J1643 and to study a possible link with the circular polarisation period, we constructed radial velocity and equivalent width (EW) curves which were subjected to a period analysis. We will focus on the July 2003 spectropolarimetric data, as J1643 could have been on a different accretion state during the 2002 observations. Nevertheless, we also analysed the 2002 radial velocity and EW curves for comparison. 

The radial velocities of the emission lines were obtained by cross-correlation of the line profile with a Gaussian template having a FWHM similar to the line width. This technique provided better results (i.e. stronger main peaks in the periodograms) than the double-Gaussian approach of \cite{schne-young}.

\subsection[]{\hel{ii}{4686} and the Bowen blend}

In Fig.~\ref{fig:rvel_ew_scargle_2003} we show the \hel{ii}{4686}, Bowen blend, \Ha, and \Hb~Scargle periodograms computed from the 2003 radial velocity and EW curves. 

The \hel{ii}{4686} power spectra do not exhibit a peak at the circular polarisation frequency ($\omega_\mathrm{pol}$). Instead, the radial velocities present the strongest peak ($> 3\sigma$) at $65.9 \pm 2.0~\mathrm{d}^{-1}$, which we identify with $\omega_\mathrm{pol}-\Omega$, where $\Omega= 8.3~\mathrm{d}^{-1}$ is the orbital frequency. The Scargle periodogram of the \hel{ii}{4686} EW curve is dominated by a peak centred at $\omega_\mathrm{EW} = 36.7 \pm 2.0~\mathrm{d}^{-1}$, which has a significance much in excess of the $4\sigma$ level. The fact that its frequency is almost exactly $\omega_\mathrm{pol}/2$ indicates that the circular polarisation is not modulated at the fundamental frequency, but the first harmonic of $\omega_\mathrm{EW}$. This will we discussed in Sect.~\ref{discuss:periods}.

The periodogram of the Bowen-blend velocities exhibits a peak very close to $\omega_\mathrm{pol}$ exceeding $2\sigma$ but not reaching the $3\sigma$ level. Unlike \hel{ii}{4686}, no significant power is observed at either $\omega_\mathrm{pol}/2$ or $\omega_\mathrm{pol}-\Omega$. However, the Bowen-blend EW curve produces a significant peak at $\omega_\mathrm{pol}/2$ but with less than half the power of the same peak in the \hel{ii}{4686} EW periodogram.

\subsection[]{Balmer lines}

Both the Balmer radial velocity and EW curves are dominated by the orbital motion. The strongest peak in the \Ha~and \Hb~periodograms (\Hg~showed the same features and is not shown) corresponds to twice the orbital frequency ($2\Omega$), and the second strongest is found at $\omega_\mathrm{pol}/2$. Note that the EW periodograms were computed from the EW curves after removing the orbital modulation to illustrate the effect of detrending.

\begin{figure*}
\includegraphics[width=85mm]{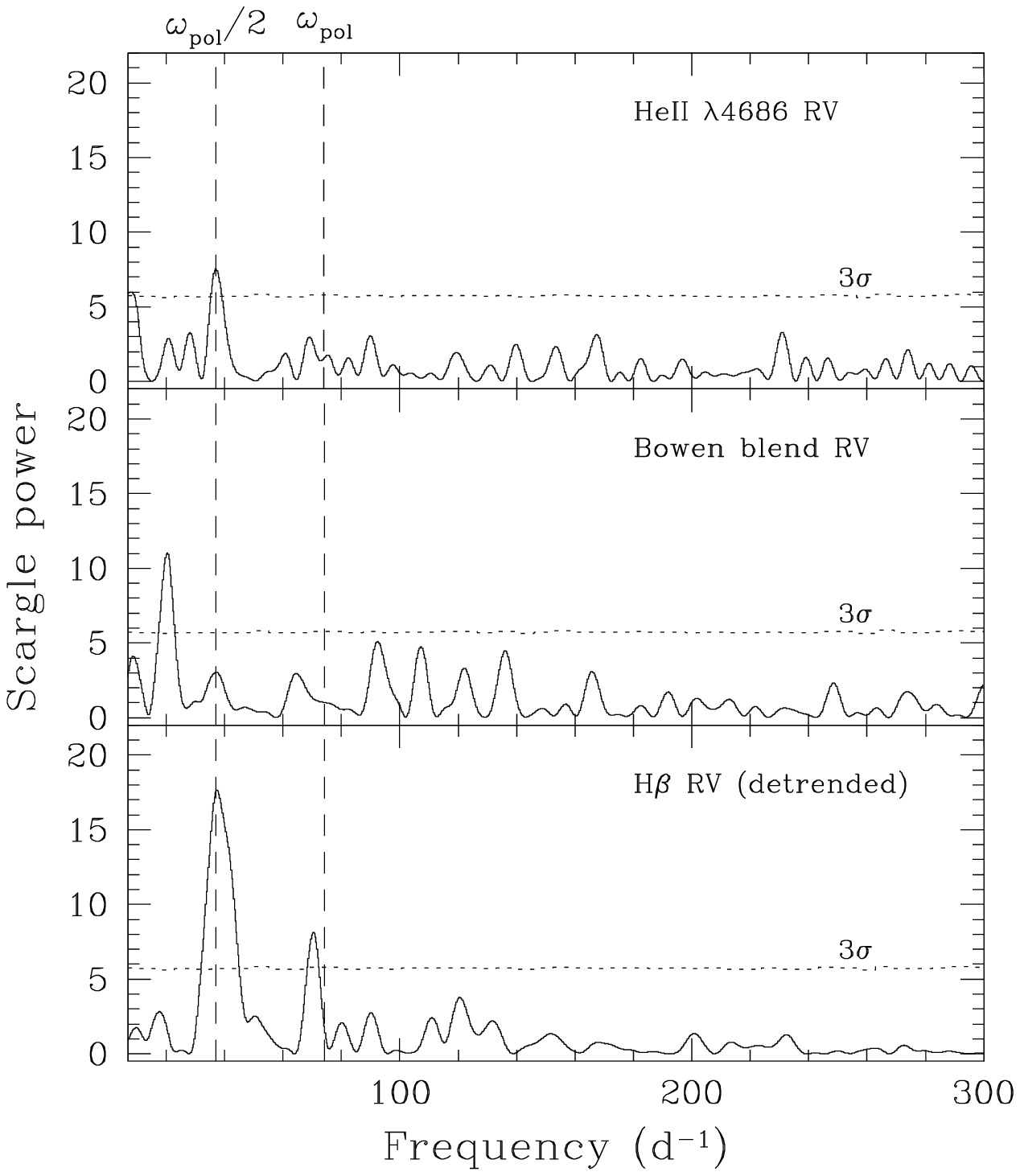} \includegraphics[width=85mm]{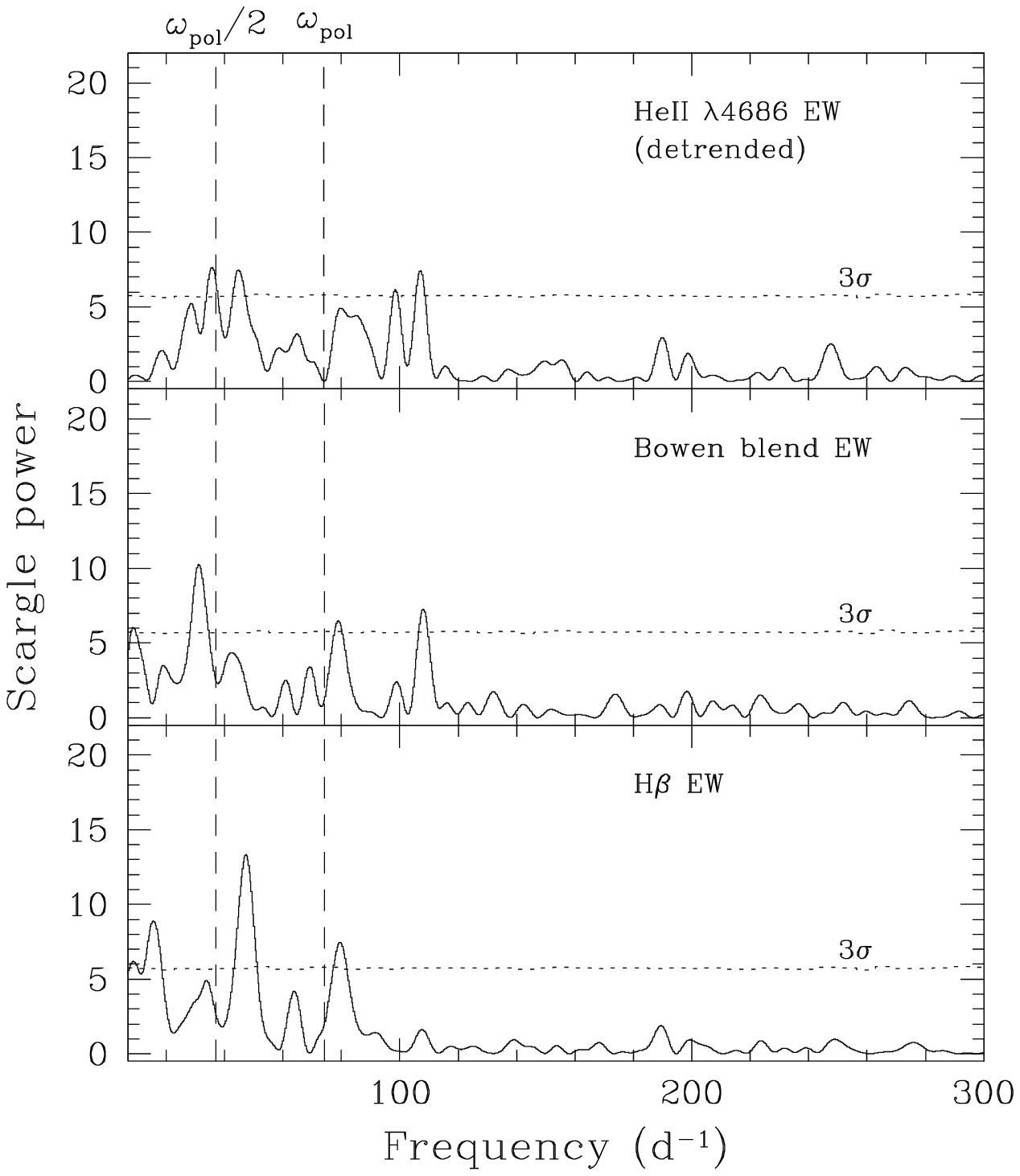}
\caption{{\sl Left panel}: Scargle periodograms of the 2002 \hel{ii}{4686}, Bowen blend, and \Hb~radial velocity curves. {\sl Right panel}: same for the EW curves.}
\label{fig:rvel_ew_scargle_2002}
\end{figure*}

\subsection[]{The 2002 data periodograms}

We have performed the same period analysis on the 2002 pure spectroscopic data. The Scargle periodograms are presented in Fig.~\ref{fig:rvel_ew_scargle_2002}. It is clear from the figure that the 2002 data portrait a different scenario. The \hel{ii}{4686} radial velocities also show a significant peak at $\omega_\mathrm{pol}/2$, but the strongest peak seen in 2003 at $\omega_\mathrm{pol} - \Omega$ is absent. The $\omega_\mathrm{pol}/2$ frequency is also the dominant one in the Balmer radial velocity periodograms (only \Hb~is shown in Fig.~\ref{fig:rvel_ew_scargle_2002}) after subtracting the original orbital modulation. The Balmer velocities also show a peak very close to $\omega_\mathrm{pol}$ at $70.4 \pm 3.0~\mathrm{d}^{-1}$. With regard to the Bowen blend velocities, a peak at $20.2 \pm 3.0~\mathrm{d}^{-1}$ shows power in excess of $3\sigma$.

The \hel{ii}{4686} EWs are dominated by a modulation at $2\Omega$. The periodogram shown in Fig.~\ref{fig:rvel_ew_scargle_2002} was constructed after removing the orbital modulation. The 2002 periodogram shows a cluster of peaks centred at $36.3 \pm 3.0~\mathrm{d}^{-1}$, which we identify as $\omega_\mathrm{pol}/2$. The flanking peaks are located at $28.2 \pm 3.3$ and $44.9 \pm 3.4~\mathrm{d}^{-1}$. Their respective distances to the central one at $\omega_\mathrm{pol}/2$ are consistent with the orbital frequency ($\Omega$). On the other hand, a peak at $31~\mathrm{d}^{-1}$ dominates the Bowen blend EWs, while one at $47.3 \pm 3.5~\mathrm{d}^{-1}$ is the strongest observed in the Balmer EW periodograms. These are consistent, within the errors, with the peaks observed in the \hel{ii}{4686} EW periodogram.

The only frequency clearly common to both the 2002 and 2003 data sets is $\omega_\mathrm{pol}/2$. The remaining (i.e. $\sim 20$, $\sim 28$, and $\sim 45~\mathrm{d}^{-1}$) are not significant frequencies in the 2003 circular polarisation data. Perhaps the most intuitive association would be $\sim 28$ and $\sim 45~\mathrm{d}^{-1}$ with $(\omega_\mathrm{pol}/2 - \Omega)$ and $(\omega_\mathrm{pol}/2 + \Omega)$, respectively. Finally, we found no explanation for the $\sim 20~\mathrm{d}^{-1}$ frequency observed in the Bowen radial velocities.






\section[]{Discussion}

\subsection{The magnetic nature of RX\,J1643.7+3402 \label{discuss:periods}}

Several authors have provided clues of the presence of magnetic white dwarfs in the SW\,Sex stars (see e.g. \citealt{casares96}, \citealt{mpais99}, \citealt{prguez2002a} or \citealt{patterson-etal2002}). Direct confirmation of magnetism in a particular system would require at least one of the two following observational facts: 1) the presence of coherent pulsations related to the 
white dwarf spin period and/or the orbital period and, mainly, 2) the detection of variable circular polarisation. The observation of modulated circular polarisation is an unambiguous proof as it implies the direct detection of a relatively strong magnetic field. It is clear from our results that the presence of an asynchronously rotating magnetic white dwarf in J1643 is beyond any doubt: the detection of variable circular polarisation and emission-line flaring (which introduces oscillations in both the EWs and the radial velocities) are solid proofs. J1643 is therefore the second SW\,Sex star in which a magnetic white dwarf has been discovered by means of spectropolarimetric techniques.




\subsection{The observed periodicities in the magnetic scenario}

Our discovery of the first SW\,Sex star showing modulated circular polarisation, LS Peg, led us to propose a magnetic model (P01) in which the periodicities 
found in the present work can be interpreted. Briefly, due to the high accretion rate, the gas stream coming from the secondary star overflows the disc and hits the magnetosphere of the white dwarf, which is assumed to extend up to the corrotation radius. From this point on, the gas couples to the magnetic field lines, eventually shocking against the white dwarf surface and forming accretion columns above the magnetic poles. The resulting X-ray radiation irradiates other structures of the system which can reprocess the high energy radiation into optical wavelengths. For example, in the intermediate polar CVs periodicities at a frequency $\omega-\Omega$ (with $\omega$ the white dwarf spin frequency) are usually detected due to the irradiation of structures anchored to the orbit (disc, secondary star, etc; \citealt{warner86}).  

According to our magnetic model, and by analogy with LS\,Peg and intermediate polars, the emission lines should be flaring at $\omega - \Omega$ as the white dwarf magnetosphere is connected to the overflown gas stream which provides information about the orbital motion. Therefore, we would identify $\omega_\mathrm{EW} = 36.7 \pm 2.0~\mathrm{d}^{-1}$ with $\omega - \Omega$. However, in this scenario the actual circular polarisation would be modulated at $2(\omega-\Omega)$. This has two problems: first, the circular polarisation is not modulated at a fundamental frequency, but at a first harmonic; and second, its frequency is related to $\omega-\Omega$, instead of $\omega$.

Modulations at the first harmonic of the spin frequency have been observed in the {\it bona fide} intermediate polar CV, PQ\,Gem. In this system, the relative dominance of the first harmonic with respect to the fundamental frequency depends on the wavelength range. In the X-ray regime, \cite{masonetal92} observed a dip at medium energies ($2-6$ keV) in their {\it Ginga} data, which produced a double-humped light curve when folded on the spin frequency, $\omega$. A periodogram of those data would therefore have provided the strongest peak at $2\omega$. In optical wavelengths, \cite{potteretal97-1} noticed that both the circular polarisation and the light curves were dominated by oscillations at $2\omega$ towards red wavelengths. The authors explain the double-waved $I$-band circular polarisation curve as a dip in the negative circular polarisation due to absorption or scattering of the cyclotron emission by the gas stream. On the other hand, the polar CV, V1432\,Aql (\citealt{rana-etal2005}), shows the same phenomenology (i.e. polarisation modulated at twice the white dwarf spin frequency). It is worth mentioning that this system is not synchronized, so it constitutes a valuable example to explain the evolution of magnetic CVs. \citeauthor{rana-etal2005} interpret this periodicity as due to the double-peaked shape of the white dwarf spin-modulated curve. Overall, these findings are not unexpected as cyclotron radiation dominates in the red and, therefore, interaction with the accretion structures can make this kind of behaviour more noticeable.

The second problem is the most fundamental question: is $\omega_\mathrm{EW}=\omega_\mathrm{pol}/2$ in J1643 the actual spin frequency ($\omega$) or the beat frequency ($\omega-\Omega$)? With the data at hand and the lack of X-ray and ultraviolet light curves, and circular polarisation data in 2002, answering to this question proves very difficult. Although we can not rule ot a disc-fed configuration (i.e. $\omega$), the circular polarisation would be modulated at the beat period with two maxima and two minima per cycle, i.e. at $2(\omega - \Omega)$, if our stream-fed model holds. In this regard, it is widely accepted that both the X-ray emission and the circular polarisation in intermediate polars show variation at the spin frequency (or its first harmonic as we discussed above). But \cite{buckleyetal97} observed the X-ray emission of the intermediate polar V2400\,Oph modulated at $\omega-\Omega$ as a result of variable accretion rate with the rotation of the magnetic field. In our SW\,Sex magnetic model, accretion along the magnetic field lines on to the white dwarf surface only takes place when the field encounters the gas stream. Therefore, optically-thick conditions for cyclotron radiation emission are only met every synodic period as seen by the gas stream. As a result, the circular polarisation can be varying at $\omega - \Omega$ or its harmonics.

Assuming a one solar-mass white dwarf, the corotation radius (with $P_\mathrm{spin}=1900$ s) is $\sim 2.3\times 10^5$ km, and the corresponding free-fall velocity is $\sim 1000$ km\,s$^{-1}$. Therefore, the timescale to refill the portion of the stream sweeped by the magnetic field is $\sim 200$ s, short enough to ensure that during the time elapsed between two consecutive passes of the field by the stream ($\sim$ 1000 s), it can be actually refilled. 

In the described accretion scenario, the white dwarf spin frequency would then be $\sim 45~\mathrm{d}^{-1}$. Peaks very close to this frequency ($> 3\sigma$) are observed in both the \hel{ii}{4686} and Balmer periodograms constructed from the 2002 EW curves, but no sign of this frequency is found in the 2003 spectropolarimetric data. The different behaviour seen in the 2002 and 2003 data likely points to changes in the accretion mode in J1643, and reflects the importance of spectropolarimetry: circular polarisation and emission-line variability can be compared as they are obtained simultaneously. Imaging polarimetry can be the only way to measure fainter objects, but in that case, time-resolved spectroscopy has to be performed as close in time as possible to avoid data contamination by accretion mode changes.

\section{Conclusions}

We have found significant circular polarisation levels modulated at a period of 19.4 ($\pm0.4$) min, which we interpret as half the beat period between the white dwarf spin period and the orbital period. In addition, the variation of the EWs and radial velocities of the \hel{ii}{4686} and Balmer lines show a predominant period at the beat period. All this phenomenology confirms the presence of a magnetic white dwarf in J1643, a characteristic that may be common to the SW\,Sex stars.

The interpretation of the observed frequencies was based on our magnetic model of an overflowing gas stream striking the white dwarf's magnetosphere at 
approximately the corrotation radius, inside which the material accretes along the magnetic field lines. This model was also proposed by us for LS\,Peg, another SW\,Sex star in which emission-line flaring and variable circular polarisation were also detected. However, in the case of J1643, the frequency at which the polarisation varies can be $2(\omega-\Omega)$, whilst for LS\,Peg it is $\omega$. However, it is also possible that the polarisation in J1643 is modulated at $2\omega$, while the EWs are at $\omega$. We are not able to address an unambiguous conclusion with the data presented in this paper.




Our results demonstrate that performing polarimetric measurements of other SW\,Sex stars is crucial to test the impact of magnetism within the class. Specially interesting are non-eclipsing systems since, following our model, in eclipsing ones the disc may be intercepting the light coming from the regions 
near the white dwarf surface. However, the results will depend on the system geometry. In fact, several eclipsing systems are known to exhibit emission-line flaring (BT\,Mon, DW\,UMa) and these should also be observed using polarimetric techniques.

\section*{Acknowledgments}

We thank Pasi Hakala, the referee, for a very constructive report on the paper. Also thanks to Tom Marsh for the use of his {\sc molly} package for time-series analysis of spectra. The William Herschel telescope is operated on the island of La Palma by the Isaac Newton Group in the Spanish Observatorio del Roque de los Muchachos of the Instituto de 
Astrof\'\i sica de Canarias.

\bsp

\label{lastpage}


\begin{thebibliography}{99}

\bibitem[\protect\citeauthoryear{Baskill et al.}{2005}]{baskilletal} Baskill D. S., Wheatley P. J., Osborne P., 2005, MNRAS, 357, 626

\bibitem[\protect\citeauthoryear{Buckley et al.}{1997}]{buckleyetal97} Buckley, D. A. H., Haberl, F., Motch, C., Pollard, K., Schwarzenberg-Czerny, A., Sekiguchi, K., 1997, MNRAS, 287, 117

\bibitem[\protect\citeauthoryear{Casares et al.}{1996}]{casares96} Casares J., Mart\'\i nez-Pais I.G., Marsh T.R., 
Charles P.A., L\'azaro C., 1996, MNRAS, 278, 219

\bibitem[\protect\citeauthoryear{Dickinson et al.}{1997}]{dickinson} Dickinson R.J., Prinja R.K., Rosen S.R., King A.R., 
Hellier C., Horne K., 1997, MNRAS, 286, 447

\bibitem[\protect\citeauthoryear{Hoard \& Szkody}{1997}]{hoard97} Hoard, D.~W., \& 
Szkody, P., 1997, ApJ, 481, 433

\bibitem[\protect\citeauthoryear{Horne}{1986}]{horne86} Horne K., 1986, PASP, 98, 609

\bibitem[\protect\citeauthoryear{Marsh \& Duck}{1996}]{marsh+duck96} Marsh, T.~R, \& Duck, S.~R., 1996, New Astron., 1, 97 

\bibitem[\protect\citeauthoryear{Mart\'\i nez-Pais et al.}{1999}]{mpais99} Mart\'\i nez-Pais I.G., Rodr\'\i guez-Gil P., 
Casares J., 1999, MNRAS, 305, 661

\bibitem[\protect\citeauthoryear{Mason et al.}{1992}]{masonetal92} Mason, K. O., et al., 1992, MNRAS, 258, 749 

\bibitem[\protect\citeauthoryear{Mickaelian et al.}{2002}]{micka} Mickaelian A. M., Balayan S. K., Ilovaisky S. A., Chevalier C., V\'eron-Cetty M.-P., V\'eron P., 2002, A\&A, 381, 894

\bibitem[\protect\citeauthoryear{Patterson et al.}{1992}]{patterson-etal92} Patterson J., Schwartz D.A., Pye J.P., Blair W.P., Williams G.A.,
Caillault J.-P., 1992, ApJ, 392, 233

\bibitem[\protect\citeauthoryear{Patterson et al.}{2002}]{patterson-etal2002} Patterson J. et al., 2002, PASP, 114, 1364

\bibitem[\protect\citeauthoryear{Piirola, Hakala \& Coyne}{1993}]{piirolaetal93-1} {Piirola} V., {Hakala} P., {Coyne}, G.~V., 1993, ApJL, 410, 107

\bibitem[\protect\citeauthoryear{Potter et al.}{1997}]{potteretal97-1} Potter S. B., Cropper M., Mason K. O., Hough J. H., Bailey J. A., 1997, MNRAS, 285, 82

\bibitem[\protect\citeauthoryear{Rana et al.}{2005}]{rana-etal2005} Rana V.R., Singh K.P., Barret P.E., Buckley D.A.H., 2005, ApJ, 625, 351

\bibitem[\protect\citeauthoryear{Rodr\'\i guez-Gil et al.}{2001}]{prguez2001} Rodr\'\i guez-Gil P.,
Casares J., Mart\'\i nez-Pais I.G., Hakala P., Steeghs D., 2001, ApJL, 548, 49

\bibitem[\protect\citeauthoryear{Rodr\'\i guez-Gil \& Mart\'\i nez-Pais}{2002}]{prguez2002a} Rodr\'\i guez-Gil P., Mart\'\i nez-Pais I.G., 
2002, MNRAS, 337,209

\bibitem[\protect\citeauthoryear{Rodr\'\i guez-Gil et al.}{2002}]{prguez2002b} Rodr\'\i guez-Gil P., Casares J., Mart\'\i nez-Pais I.G., 
Hakala P., 2002, in Proc. The Physics of Cataclysmic Variables and Related Objects. Gottingen, Germany (astro-ph/0110173)

\bibitem[\protect\citeauthoryear{Rodr{\'{\i}}guez-Gil et al.}{2004}]{prguez2004} 
Rodr{\'{\i}}guez-Gil, P., G{\"a}nsicke, B.~T., Barwig, H., Hagen, H.-J., \& 
Engels, D.\ 2004, A\&A, 424, 647

\bibitem[\protect\citeauthoryear{Rodr{\'{\i}}guez-Gil, Schmidtobreick \& G\"ansicke}{2007}]{prguez2007a} Rodr\'\i guez-Gil P., Schmidtobreick
L., G\"ansicke B.T., 2007, MNRAS, 374, 1359

\bibitem[\protect\citeauthoryear{Rodr{\'{\i}}guez-Gil et al.}{2007}]{prguez2007b} Rodr\'\i guez-Gil et al., 2007, MNRAS, 377, 1747

\bibitem[\protect\citeauthoryear{Rosen, Mittaz \& Hakala}{1993}]{rosenetal93-1} Rosen S. R., Mittaz J. P. D., Hakala P. J., 1993, MNRAS, 264, 171

\bibitem[\protect\citeauthoryear{Scargle}{1982}]{scargle} Scargle J.D., 1982, ApJ, 263, 835

\bibitem[\protect\citeauthoryear{Schneider \& Young}{1980}]{schne-young} Schneider D.P., Young P., 1980, ApJ, 238, 946

\bibitem[\protect\citeauthoryear{Simic et al.}{1998}]{simic} Simic, D., Barwig, H., 
Bobinger, A., Mantel, K.-H., \& Wolf, S.\ 1998, A\&A, 329, 115

\bibitem[\protect\citeauthoryear{Smith, Dhillon \& Marsh}{1998}]{smi-dhi-marsh} Smith D.A., Dhillon V.S., Marsh T.R., 1998, MNRAS, 296, 465


\bibitem[\protect\citeauthoryear{Szkody \& Piche}{1990}]{szko-pi} Szkody P., Pich\'e F., 1990, ApJ, 361, 235


\bibitem[\protect\citeauthoryear{Thorstensen et al.}{1991}]{thorstensen} Thorstensen J.R., Ringwald F.A., Wade R.A., Schmidt G.D., Norsworthy
J.E., 1991, AJ, 102, 272

\bibitem[\protect\citeauthoryear{Tinbergen \& Rutten}{1992}]{tin-rut} Tinbergen J., Rutten R.G.M., 1992, 
User Manual 21, Isaac Newton Group, La Palma

\bibitem[\protect\citeauthoryear{Turnshek et al.}{1990}]{turnshek} Turnshek D.A., Bohlin R.C., Williamson II R.L., Lupie O.L., Koorneef J., 
1990, AJ, 99, 1243

\bibitem[\protect\citeauthoryear{Warner}{1986}]{warner86} Warner B., 1986, MNRAS, 219, 347

\bibitem[\protect\citeauthoryear{Williams}{1989}]{williams89} Williams R.E., 1989, AJ, 97, 1752





\end{thebibliography}
\end{document}